# The end of multiple choice tests: using AI to enhance assessment.

Running title: Using AI to enhance multiple choice test

**Michael.W. Klymkowsky** [1*] **& Melanie M. Cooper** [2]

[1] Molecular, Cellular, and Developmental Biology & CU Teach
 University of Colorado Boulder, Boulder, CO. 80309  ORCID:0000-0001-5816-9771
[2] Chemistry, Michigan State University, East Lansing, MI. 48824  ORCID:0000-0002-7050-8649

* Corresponding Author: MCD Biology, 1945 Colorado Avenue University of
 Colorado Boulder, Boulder, CO 80309
 T: 720-425-8266 email: klym@colorado.edu

**Disclosure**: This work was carried out using the CustomGPT.ai system; we have no financial relationship with CustomGPT.ai, although we have received favorable terms for its use, and MK was recently awarded a "AI Research Grant for College Faculty" from CustomGPT.ai.

**Nota bene:** Echoing Thibaut Brunet (on Twitter/X): We are not planning to submit the story to a journal right away - instead, we see this as a "living" preprint that we'll update based on feedback from the community and new data. Comments/suggestions/criticisms are all very welcome.

**Abstract**:

Effective teaching relies on knowing what students know-or think they know. Revealing student thinking is challenging. Often used because of their ease of grading, even the best multiple choice tests, those using research based distractors (wrong answers) are intrinsically limited in the insights they provide due to two factors. When distractors do not reflect student beliefs they can be ignored, increasing the likelihood that the correct answer will be chosen by chance. Moreover, making the correct choice does not guarantee that the student understands why it is correct. To address these limitations, we recommend asking students to explain why "wrong" choices are wrong and why they chose their answer. Using a discipline-trained artificial intelligence-based bot it is possible to analyze their explanations, identifying the concepts and scientific principles that maybe missing or misapplied. The bot also makes suggestions for how instructors can use these data to better guide student thinking. In a small "proof of concept" study, we tested this approach using questions from the Biology Concepts Instrument (BCI). The result was rapid, informative, and provided actionable feedback on student thinking. It appears that the use of AI can address the weaknesses of conventional multiple choice test. It seems likely that incorporating AI-analyzed formative assessments will lead to improved overall learning outcomes.

**Introduction**

      Examining and acting on student responses to assessment questions is critical for keeping instructors "in touch" with how students are responding to, using, and integrating (or not) the information and challenges they are presented with (see Shepard, 2000, 2021 and references therein).  Students' prior knowledge, assumptions, and skills impact how they respond to learning materials. Yet "assessment and instruction are often conceived as curiously separate in both time and purpose" (Graue 1993, quoted in Shepard 2000). To discover how new, often counter-intuitive scientific ideas or unexpected facts (empirical observations) interact with students' implicit, often unrecognized assumptions requires that we give students opportunities to explain their thinking.  While one-on-one and small group discussions would  be optimal, these are rarely possible or practical, particularly in the context of the large enrollment college courses that students are often subjected to.  In lieu of such exploratory conversations, we turn to assessments of various types.  Assessments also serve, often implicitly, to inform students as to what is considered important for them to know or be able to do in a course (see McClymer & Knowles, 1992, Stowe et al., 2021). What assessments aim to reveal can vary from memorization and recognition to the ability to use knowledge of relevant facts, concepts, and scientific principles to develop and apply plausible models, mechanistic explanations and arguments, as well as the evaluation and interpretation of experimental data. Such questions probe what has come to be known as three-dimensional learning (see National Research Council, 2012). Tools such as the three-dimensional Learning Assessment Protocol (3D-LAP) can identify questions that "have the potential to elicit evidence of three-dimensional learning and those that do not" (Laverty et al., 2016).

      Assessments that elicit appropriate evidence can provide students and instructors with feedback about what a student knows, and whether instruction has been effective. In the context of a traditional, but now rare, oral exam examiners are able to interact and follow up with the student, encouraging them to reflect upon and revise their assumptions, and to consider their appropriateness (see Theobald, 2021).  In larger groups, the process of analyzing student responses, that may include drawings and text, and then providing students and instructors with actionable feedback is necessarily more difficult and time consuming. Feedback takes time to generate and deliver; It takes time and resources to establish a consensus as to what makes for an acceptable answer and there can be issues with the consistency of the analyses deliver to students. Even when thoughtful feedback can be delivered to the student, they may or may not be required to "use it", that is to respond to it and revise their answers.  Indeed, it appears that much of the feedback on summative assessments goes unread and unused (see Jonnson, 2013, Winston et al., 2017).

      The end-result is that a number of forces have encouraged the use of multiple-choice tests, particularly in large enrollment courses (see Loepp 2021). Multiple choice questions can be machine-graded, to provide rapid and consistent correct/incorrect feedback, without the interpretive complexity associated with the evaluation of written answers. At the same time, there is evidence that suggests that such test produce biased results (see Griselda, 2021), and tend to over-estimate what students understand (see Sadler, 2005, Lee et al., 2011). A further limitation of multiple-choice questions is that we cannot know exactly why a student did not choose a wrong answer, did they know why it was wrong?  It is worth noting that there are often implicit clues as to the correct answer, unrelated to their correctness (see Mentzer, 1982); these clues can be used to "pass" a multiple-choice test without actually understanding the subject of the test.[1]

---

[1] Or so we have been informed in a recent conversations with a radiotherapist.

**Research-based distractor-driven multiple-choice tests:** Most multiple-choice tests are generated either by instructors or supplied by commercial content providers as part of their "instructional packages". A major weakness of many multiple-choice tests are the wrong choices. Often these choices do not reflect ideas held by students, and are ignored, increasing the probability of choosing the correct answer by chance rather than based on an understanding of why it is correct. The work of Schneps & Sadler (1987), and their film "A Private Universe", revealed that students often hold persistent assumptions that explicitly contradict what they have been "taught". These implicit, but often unrecognized, beliefs may develop both through personal experiences and in response to previous instruction (see Sadler, 1998, Briggs et al., 2017).

Efforts to identify the various ideas influencing student thinking, through extensive one-on-one interviews, led to the identification of commonly held, so-called "misconceptions". These misconceptions could then be used as the basis of the wrong answers, known as distractors, on multiple choice tests. As noted by Sadler (1998) "This kind of instrument, in which distractors match common student ideas, has a profoundly different psychometric profile from conventional tests and exposes the weakness evident in conventional standardized tests. Distractor-driven multiple-choice tests combine the richness of qualitative research with the power of quantitative assessment, measuring conceptual change along a single uniform dimension."

Multiple choice tests in which the distractors are developed through research into student thinking have come to be known as concept inventories. Concept inventories offer an easily administered, often in a pre-post setting, approach to quantifying learning outcomes (see Evans et al., 2003; Klymkowsky et al, 2003; Garvin-Doxas et al., 2007, Lindell, et al., 2007, Libarkin, 2008, Bretz et al., 2014, Sands et al., 2018; Klymkowsky & Garvin-Doxas, 2020). That said, the research involved in the development of a concept inventory requires a significant investment and the conclusions that can be drawn from student responses are necessarily limited (discussed below).

Among the first, and arguably the most influential concept inventory is the Force Concept Inventory (FCI) developed by Hestenes et al., (1992). It has been subject to a number of analyzes by the Physics Education Research (PER) community, and it is worth noting that exactly what it measures remains problematic (see Huffman & Heller, 1995). As an example, it is rare that the "universe" of distracting ideas is completely covered in any concept test. For example when Rebello & Zollman (2004) "examined students' open-ended responses to FCI questions, their analyses revealed that common aspects of student thinking are not found in the original FCI distractors" (quoted from Klymkowsky & Garvin-Doxas, 2020). The introduction of new, presumably more distracting distractors led to lower FCI scores.

The absence of direct evidence for why student do not choose a distractor and why they choose the correct answer means that concept inventories are also limited as assessment instruments (Smith & Tanner, 2010; Wallace & Bailey, 2010, Stowe & Cooper, 2019, Klymkowsky & Garvin-Doxas, 2020, Cooper & Klymkowsky, 2022 and others). A focus on persistent misconceptions has become less prevalent; many researchers now ascribe to a more fluid approach to how students construct and use knowledge. Students have both cognitive and epistemic resources that can be woven together on the fly in response to a specific context or prompt (see Hammer & Elby, 2003, diSessa, 2014). We can, however, enhance what can be deduced about student thinking if we examine the rationale and logic students use as they choose one answer among many (Talanquer, 2017; Scott & Schumayer, 2018; Sikorski & Lee 2024).

In a recent review chapter of assessment strategies, we wrote that for concept inventories' "Future extensions will require an iterative feedback system that can analyze students' textual responses "on the fly" and pose disciplinarily relevant and clarifying Socratic questions" (Cooper & Klymkowsky 2022). Little did we anticipate that the future, in the form of generative artificial intelligence and large language models, was already upon us. Using the CustomGPT.ai system, a generative AI system that uses Retrieval Augmented Generation (RAG) (see Chen et al 2024) we developed a ChatBot trained to evaluate student responses based on various sources related to cell and molecular biology. These sources include our freely available CLUE chemistry materials (Cooper & Klymkowsky, 2013) and biofundamentals (Klymkowsky et al., 2016), an introductory molecular biology text (a full list of sources is presented in Supplement 1). In contrast to ChatGPT4, the bot does not answer questions beyond its "training set". The use of a defined training set minimizes the problem of hallucination. We have not seen hallucinogenic responses to date (unpublished observations). This ChatBot, named Dewey after the education reformer John Dewey (1859 – 1952) (see Wikipedia entry), was given the following persona "Dewey, the bot, is able to analyze student answers to a wide range of questions, providing a detailed summary of necessary facts and concepts that are missing or misapplied in student answers, as well as generating suggestions for instructional approaches to address them." Dewey and interactive ChatBots like it make possible a simple way to update conventional multiple choice tests. The bot can analyze multiple choice questions before they are presented to students. For example, the Dewey chatbot can characterize questions asked in terms of whether they require mechanistic or conceptual thinking or primarily involve the recall (memorization) of facts presented (see Supplement 2).

Following on from Sikorski & Lee (2024), we view "concept inventories as a ready-made tool teachers can adapt". Here is one simple adaptation made possible by the availability of AI bots. In the context of a formative multiple choice/concept inventory-type assessment, we ask students to explain why they chose their answer and why the other answers are wrong. Dewey is then asked to analyze those responses, and then supply the instructor with that analysis. In our tests, Dewey responded appropriately to a wide range of topics characteristic of undergraduate biology programs centered on molecular and cellular processes (data not shown). Response analysis and the generation of instructional suggestions by Dewey is essentially immediate. The result is that the instructor has the time needed to thoughtfully consider how to adapt their instructional efforts, to introduce or review background materials, and to engage students with targeted in-class questions.

In a "proof of concept" study, we asked a small group of graduate students in molecular biology to explain their choice of the correct answer and why the incorrect responses to two questions from the Biology Concepts Instrument (Klymkowsky et al., 2010, Champagne Queloz et al., 2017) were wrong. Student responses [copied verbatim and shown in Supplement 3] were analyzed (see below). The analysis of a second BCI question is presented in Supplement 4.

**Dewey was asked:** "In response to this question (How is genetic drift like molecular diffusion? The choices were
      A. Both are the result of directed movements.
      B. Both involve passing through a barrier.
      C. Both involve random events without regard to ultimate outcome.
      D. They are not alike. Genetic drift is random; diffusion typically has a direction.)

Here are their explanations [student responses removed]. Provide an analysis of students explanations and identify misconceptions and misunderstandings.

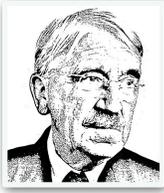

**Dewey's response (in italics):**

*Analyzing the student responses to the question about the similarities between genetic drift and molecular diffusion reveals a mix of accurate understanding and misconceptions. Here's a breakdown based on the answer choices provided:*

*Answer Choice A: Both are the result of directed movements.*

***Misconception Identified**: Many students incorrectly associate "directed movements" with purpose or intention, which neither genetic drift nor molecular diffusion inherently possess. The correct understanding is that both processes are fundamentally random; molecular diffusion is influenced by concentration gradients, but this does not imply a purposeful direction.*

***Percentage of Misconception**: Approximately 90% of the responses show a misunderstanding of the term "directed movements" in the context of these natural processes.*

***Answer Choice B**: Both involve passing through a barrier.*

***Misconception Identified**: Students often confuse the concept of barriers in molecular diffusion (e.g., semi-permeable membranes) with any form of barrier in genetic drift. Genetic drift does not involve physical barriers but is a change in allele frequencies due to random sampling.*

***Percentage of Misconception**: Around 80% of the responses incorrectly relate barriers to genetic drift, showing a significant conceptual gap.*

***Answer Choice C**: Both involve random events without regard to ultimate outcome.*

***Correct Understanding:** This was the correct answer, but some students still showed confusion, particularly regarding the outcome of diffusion (equilibrium) and its characterization as an "ultimate outcome."*

***Percentage of Correct Understanding**: Only about 10% of students correctly identified and explained why this choice was accurate without adding unnecessary or incorrect details.*

***Answer Choice D**: They are not alike. Genetic drift is random; diffusion typically has a direction.*

***Misconception Identified**: Students struggle with the concept of "direction" in diffusion. While it's true that diffusion involves movement from higher to lower concentration areas, this is a result of random molecular motion, not a directed or purposeful process. Some students mistakenly attribute a non-random, directed quality to diffusion.*

***Percentage of Misconception**: Approximately 70% of the responses misinterpret the nature of direction in diffusion, confusing it with intentional or non-random movement.*

***Suggestions for Instructional Improvement:***

***Clarify the Concept of Randomness**: Emphasize that both genetic drift and molecular diffusion are processes influenced by random events. Use visual aids or simulations to show how these processes occur without any directional intent.*

***Address Misconceptions about Barriers**: Specifically teach the differences in how barriers function in biological contexts, distinguishing between physical barriers in diffusion and metaphorical or statistical "barriers" in genetic drift.*

***Enhance Understanding of "Directed Movements"****: Use examples and counterexamples to teach what is meant by "directed" in scientific contexts, clarifying that it refers to physical influences like gradients rather than purposeful actions.*

***Overall,*** *while some students grasp the stochastic nature of these processes, there is a clear need for more targeted instruction to address prevalent misconceptions, especially concerning the concepts of direction and barriers."*

The results of this analysis can be shared with students and discussed further following the bot's instructional suggestions. As an example, we could ask follow-on (in-class) questions, to paraphrase Sikorski and Lee (2024) "What does the correct answer leave out" or "in what ways was it confusing to you, or why might a student be likely not to pick it?" If, as in the case above, where there are persistent ideas that are inappropriately applied and shared by a numbers of students, it is incumbent on the instructor to re-think how they approach these ideas and redesign their instructional activities.

**Conclusion / Summary:** It is well established that formative assessment can help instructors stay in touch with what students think and what they are learning (or not) as a course proceeds. Previously, in most cases the analysis and reporting of the results of open-ended formative assessment activities has been impractical or impossible to deliver in a timely and/or economical manner. Through the use of AI-based analyst bots, like Dewey, it becomes possible to quickly generate in-depth analyses at both the overall class level and the individual student level and deliver them to instructors. While Dewey's training set appears adequate for the questions we analyzed, it can easily be expanded to include more advanced topics as needed. In the same vein, different disciplines and degree programs will want to develop their own training libraries, suited to the learning outcomes they seek to achieve. By using short formative assessments, in the form of multiple-choice questions with student explanations for correct and incorrect answers, AI-based analyses can reveal ways of students thinking about concepts or phenomena. Such class level analyses of multiple choice/concept inventory-type questions and other formative assessments provides a richer and more readily applied evidence base about student understanding than the percent of correct answers; it includes student explanations about their thinking. What is. needed for this type of feedback to make a difference is that the idea of actionable formative assessment has to be embraced by instructors and course designers. That likely involves a serious rethinking of what is taught in a course (see Klymkowsky & Cooper, 2012, Cooper et al., 2024).

**Acknowledgements:** Thanks to Michael Shiferaw, Beiyi Xu. Clayton Lewis, and Valerie Otero for spurring us on in our thinking about and use of AI in various ways, and Nate Kocho (and CustonGPT.ai) for generously supporting the first phase of the project. The project was funded in part by funds associated with the Lappin-Phillips Endowment at MSU and "AI Research Grant for College Faculty" from CustomGPT.ai to MK.

**Conflict of interest:** The authors declare no conflict, they have no financial stake in CustomGPT.ai.

**Human Subjects:** This work was carried out under IRB Protocol 23-0389 as amended

**Supplement 1: Sources used to train Dewey, the analyst bot**

| Text | Reference |
|---|---|
| Biofundamentals.pdf (27.79 MB) | Klymkowsky, M.W., Rentsch, J.D., Begovic, E. and Cooper, M.M. (2016). The design and transformation of Biofundamentals: A nonsurvey introductory evolutionary and molecular biology course. *CBE—Life Sciences Education*, *15*(4), p.ar70. |
| CLUE-PLUS WORKING 2019.pdf (21.03 MB) | Cooper, M. and Klymkowsky, M. (2013). Chemistry, life, the universe, and everything: A new approach to general chemistry, and a model for curriculum reform. Journal of Chemical Education, 90(9), pp.1116-1122.<br>Cooper, M.M., Stowe, R.L., Crandell, O.M. and Klymkowsky, M.W. (2019). Organic chemistry, life, the universe and everything (OCLUE): A transformed organic chemistry curriculum. Journal of Chemical Education, 96(9), pp.1858-1872. |
| conceptual foundations 2010.pdf (311.19 KB) | Klymkowsky, M.W. (2010). Thinking about the conceptual foundations of the biological sciences. CBE—Life Sciences Education, 9(4), pp.405-407. |
| makin mechanistic sense.pdf (742.21 KB) | Klymkowsky, M.W. (2021). Making mechanistic sense: are we teaching students what they need to know?. Developmental Biology, 476, pp.308-313. |
| harrison-et-al-2005-stochastic-models-of-neuronal-dynamics.pdf (350.61 KB) | Harrison, L.M., David, O. and Friston, K.J. (2005). Stochastic models of neuronal dynamics. *Philosophical Transactions of the Royal Society B: Biological Sciences*, *360*(1457), pp.1075-1091. |
| Mullers Morphs.pdf (827.36 KB) | Henson, K., Cooper, M.M. and Klymkowsky, M.W. (2012). Turning randomness into meaning at the molecular level using Muller's morphs. Biology Open, 1(4), pp.405-410. |

**Supplement 2: Supplement**

**Dewey's analysis of questions Part 1 Please analyze these questions (1-29)(from Hennessey, K.M. and Freeman, S., 2023. Nationally endorsed learning objectives to improve course design in introductory biology. bioRxiv, pp.2023-10.) in terms of whether they are conceptual, mechanistic, or primarily recall. The questions labeled "A" are lower-order cognitive skill while questions labeled "B" are higher order.**

The questions provided can be categorized based on their cognitive demand—whether they require recall, understanding of concepts, or mechanistic reasoning. Here's an analysis of each question type:

1. **Lower-order cognitive skill questions (labeled "A")**:

    - These questions generally require recall or basic understanding. For example:
        - **1A, 3A, 7A, 8A, 10A, 11A, 12A, 13A, 14A, 17A, 21A, 22A, 23A, 24A, 25A, 26A, 28A**: These involve defining terms, describing structures or processes, and drawing or labeling diagrams. They are primarily focused on recalling information or applying basic concepts in straightforward ways.

2. **Higher-order cognitive skill questions (labeled "B")**:

    - These questions require deeper understanding, analysis, and application of concepts in new contexts. They often involve prediction, explanation, and the application of mechanistic reasoning. For example:
        - **2B, 4B, 5B, 8B, 9B, 10B, 13B, 15B, 16B, 18B, 19B, 20B, 21B, 22B, 23B, 25B, 27B, 29B**: These questions ask students to predict outcomes based on structural or functional knowledge, explain relationships between concepts, and apply principles to novel situations. They encourage students to think about how mechanisms work and how different factors interact.

3. **Conceptual versus Mechanistic**:

    - **Conceptual Questions**: These often involve understanding and applying overarching principles or theories to specific situations. Examples include explaining the relationship between hydrogen bonding and environmental phenomena (2B), or analyzing how the structure of biological molecules impacts their function (15B).
    - **Mechanistic Questions**: These require an understanding of the underlying mechanisms that explain how processes occur at a detailed level. For instance, predicting how changes in membrane composition affect fluidity and permeability (10B), or explaining the mechanistic role of transport proteins in membrane crossing (13B).

4. **Recall Questions**:

    - Primarily focused on retrieving factual information or describing well-known processes without requiring deep understanding or application. Examples include defining terms like acid, base, and pH (3A), or describing the functions of proteins in cells (7A).

In summary, the "A" questions are generally more focused on recall and basic application, suitable for testing foundational knowledge. The "B" questions, however, push students towards higher-order thinking, requiring them to apply concepts in new ways, predict outcomes, and understand mechanisms at a deeper level. This distinction is crucial for assessing both the breadth and depth of students' understanding in biology.

**Supplement 3: Student responses to Question 1**

In response to this question ("How is genetic drift like molecular diffusion? The choices were
A. Both are the result of directed movements.
B. Both involve passing through a barrier.
C. Both involve random events without regard to ultimate outcome.
D. They are not alike. Genetic drift is random; diffusion typically has a direction.)
Students were asked to explain why the wrong answer choices were wrong. Here are their explanations: Provide an analysis of students explanations and identify misconceptions and misunderstandings.

(A. Wrong, directed movements seems to imply a purpose, which neither of these phenomena truly have unless it is assigned to them; Wrong, genetic drift is a transgenerational effect where physical barriers are a variable as in the instance of diffusion across a membrane. Also diffusion isnt always across a barrier either; Wrong, genetic drift is indeed random, although diffusion can random. Random at the molecular level, but one could create a direction per the concentration arrangements; Incorrect. Genetic drift is the random change in the frequency of alleles from generation to generation due to random events involving the number of offspring that survive, it is not directed. Diffusion is the movement of molecules from an area of high concentration to an area of lower concentration. While it could be argued that diffusion is "directed" by circumstance, it is the result of random movement of particles (or energy) in a manner that distributes them relatively evenly throughout a solvent. At the molecular level, the randomness of movement precludes any notion of "directed" movement; wrong. Genetice drift occurs due to random events; wrong Genetic drift is not a directed movement; wrong One copy of the gene can make some protein; correct; wrong Incorrect: Although some molecular components may or may not randomly direct the movements in molecular diffusion, nothing directly plays a role in directing movements in genetic drift as genetic drift has no direction and is random; wrong Neither are directed. Genetic drift is the random change in allele frequency due to random chance and molecular diffusion is the result of the random brownian motion of molecules; wrong This is incorrect because neither genetic drift nor diffusion are directed; they both include randomness in either a mutation rate or the kinetic movement of particles; wrong Neither genetic drift nor molecular diffusion are directed movements. Genetic drift is the random change in allele frequencies due to chance and not due an evolutionary advantage. The frequencies often change over generations. Diffusion is also not a directed movement, it is the random movement of molecules that leads to molecules in high concentration area moving to an area of low concentration and becoming an even mixture)

(B. Wrong, genetic drift is a transgenerational effect where physical barriers are a variable as in the instance of diffusion across a membrane. Also diffusion isnt always across a barrier either; Incorrect. Diffusion can occur through a barrier which allows the passage of the molecule in question but this is not required. I'm actually at a loss as to how genetic drift would be related to "passing through" a barrier; wrong a barrier is not neccasary for either. I am not sure what "barrier" would be referencing; wrong Genetic drift doesn't pass through a barrier; Wrong There is no physical barrier for the population affected by genetic drift like there is in molecular diffusion; wrong Incorrect: Molecular diffusion involves molecular components passing through barriers whereas genetic drift can happen for a number of different reasons that may or may not have to do with barrier passaging; Wrong Genetic drift doesn't have a barrier, it just refers to the change in allele frequencey. Diffusion doesn't necessarily have a barrier, it could refer to a group of molecules "spreading out" in space due to brownian motion; wrong This is incorrect because neither genetic drift nor diffusion are directed; they both include randomness in either a mutation rate or the kinetic movement of particles; wrong Neither genetic drift nor molecular diffusion have to go through/over/pass a barrier for the drift/diffusion to occur)

(C - the correct answer: Incorrect. While both involve random events and genetic drift occurs with no consideration of ultimate outcome, diffusion will result in a spreading of a molecule from an area of high concentration to areas of low concentration until equilibrium is achieved. I would argue that equilibrium is the "ultimate outcome" to be achieved; wrong molecular diffusion will flow down a gradient from high to low concentration; wrong Delete the gene from one of the two chromosomes, the gene is still dominant; wrong These events are not random it is at the mercy of the conditions present at the time. For example diffusion is based on concentrations at the time and that due to human intervention in nature has brought a larger prevalence of natural disasters, habitat loss, etc; Correct: Molecular diffusion may have some directive components such as chaperone proteins that may guide diffusion however diffusion; for the most part involves random movements of molecules that may or may not diffuse and really there is no ultimate outcome for the diffusion other than random molecular "stuff" happened to be at the right place at the right time for diffusion, which is similar to genetic drift which may involve one or many components that can all be random and has no clear direction, coming together to form a genetic drift; wrong It could seem like diffusion has a direction going from high to low concentration, but each individual molecule doesn't have a specific direction, it's just random)

(D: Wrong, genetic drift is indeed random, although diffusion can random. Random at the molecular level, but one could create a direction per the concentration arrangements; Correct. While it is perhaps not immediately obvious upon first glance at a system, diffusion will occur "towards" areas of low concentration and therefore has a direction. It is however worth noting that it can have multiple directions depending on circumstance; wrong The movement of molecular during diffusion is random.; wrong One copy of the gene still can make some the transcription factor; wrong Genetic drift has been known as random but due to increasing industrializaition and commercialism increases the prevalence of natural disasters more common. Thus human presence in nature has made genetic drift more directed (Ex: habitat fragmentation causes genetic drift, and the resulting population is smaller and more susceptible to genetic drift); wrong Incorrect: Although molecular diffusion may have some directive forces associated with it, the process is ulimately random as all the movements of molecules is random and has no real direction or motive. Genetic drift is similar in that there is no direction, random events can lead to genetic drift; wrong This is incorrect because they are alike and diffusion doesn't need to have a direction-- it occurs due to the random movement of particles and that movement continues to occur without a concentration gradient;  wrong Genetic drift is random and can vary between generations. While diffusion typically goes from a high to low concentration, the individual movement of molecules is still random or by chance so I wouldn't say it has a direction. It's more that the molecules end up being homogenized due to random movement)

**Supplement 4: Second BCI Question analysis**

**In response to this question "Consider a diploid organism that is homozygous for a particular gene. How might the deletion of this gene from one of the two chromosomes produce a phenotype?**
A. If the gene encodes a multifunctional protein.
B. If one copy of the gene did not produce enough gene product.
C. If the deleted allele was dominant.
D. If the gene encoded a transcription factor"

Students were asked to explain why the wrong answer choices were wrong. Here are their explanations: Provide an analysis of students explanations and identify misconceptions and misunderstandings.

Student responses ↓

(A: Wrong, it could lead to some effects, but any effect noticed would be more related to gene dose rather than the genes ontology that permits it's multifunctionality; Incorrect. If the two genes are the same then the protein should retain all functions if one gene is deleted; wrong, It doesn't matter what protein the gene encodes if one copy of the gene doesn't produce enough gene product anyway; wrong this could be true but only if the assumption that the gene product from one allele is not efficient, but is not itself a explantation as to why; wrong If the other copy of the gene is sufficient, it shouldn't matter that it is multifunctional; wrong The other copy of the gene would also be multifunctional, likely not resulting in a different phenotype; wrong This is incorrect because even if the gene encodes for a multifunctional protein, the loss of one allele would mean that the other remaining functional allele can produce that multifunctional protein which fulfills its various roles; wrong A is wrong because the organism is diploid, so, if the protein is multifunctional, the organism still has one copy left that can produce the protein)

(B - the correct answer: Incorrect. This may affect expressivity but it will not change the phenotype;

(C: Wrong, if the allele was dominant, it only needs one copy to sustain the phenotype in most cases; Correct. A dominant allele is responsible for the expression of a phenotype. If the dominant allele is deleted then the dominant phenotype would cease to be expressed; Wrong "Dominant" just describes the allele that "masks" the other after expression. Both alleles are expressed; that is, there would be a combination of proteins resulting from the dominant AND recessive alleles. You'd get a phenotype either way; wrong the organisms is homozygous for the allele so both alleles would either be dominant or recessive; wrong If the organism is homozygous for the gene, then one wouldn't be dominant over the other, they should be the same'; wrong If the organism was homozygous, then a copy of the dominant allele is still present, resulting in the dominant (same) phenotype; This is incorrect because the organism is homozygous, meaning that both alleles lost are the same and therefore one cannot be dominant; wrong because it doesn't matter if the deleted allele is dominant or recessive, it is homozygous for the gene and so both copies would be dominant or recessive and so deleting one would not give a phenotype unless you need both copies being expressed which still doesn't have anything to do with dominant or recessive)

(D: Wrong, the genes job isn't as important as product dose sufficiency; Incorrect. Like B, this may change expressivity but it will not change the phenotype; Wrong Again, it doesn't matter what protein the gene encodes if one copy of the gene doesn't produce enough gene product anyway; wrong this could be true but only if the assumption that the gene product from one allele is not efficient, but is not itself a explantation as to why; wrong Again, assuming the other gene is producing enough of the transcription factor, this shouldn't matter because there are two copies of the gene, and the other copy will just be used instead; wrong Due to the downstream amplification of the activity of a transcription factor, the presence of the intact gene will likely result in a similar phenotype (acting in a dominant fashion); This is incorrect because one single copy of the allele could still produce the transcription factor protein that is needed to trigger specific gene expression, and therefore the obsevable phenotype has nothing to do with wether that gene is a transcription factor; wrong because even though transcription factors are crucial for gene expression and would likely create a phenotype, the question asks about specifically deleting one copy of the gene. So, again, the organism still has one copy of the gene that can be expressed)

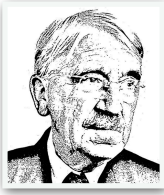 **Dewey's Analysis:** Analyzing the students' explanations for why certain answer choices are incorrect reveals a mix of accurate understanding and misconceptions about genetic principles, particularly concerning gene dosage, dominance, and the functional roles of genes.

**Misconceptions Identified:**

1. **Multifunctionality of Proteins (A):** Several students incorrectly assume that the multifunctionality of a protein is irrelevant if one gene copy remains. This overlooks scenarios where gene dosage (the amount of gene product produced) is critical, even for multifunctional proteins. The misconception here is the failure to recognize that the loss of one copy can still impact the phenotype if the remaining gene does not produce sufficient product to meet cellular demands.
2. **Gene Dosage and Expressivity (B):** The correct answer was identified, but explanations often confuse expressivity with phenotype alteration. Some students correctly note that reduced gene dosage might affect expressivity (variation in phenotype among individuals with the same genotype) but incorrectly conclude that it cannot change the phenotype outright. This is a misunderstanding, as reduced dosage can indeed lead to a phenotype if the remaining copy does not suffice for normal function.
3. **Dominance in Homozygous Alleles (C):** There is confusion about dominance in a homozygous context. Several students correctly state that in a homozygous organism, the concept of one allele being dominant over the other is irrelevant. However, some misunderstandings persist about how dominance influences phenotype expression when only one allele is present, which could be critical if the dominant allele is haploinsufficient.
4. **Role of Transcription Factors (D):** Misunderstandings here involve the role and necessity of transcription factors. Some students correctly identify that losing one copy of a gene encoding a transcription factor might not alter the phenotype if the remaining copy compensates adequately. However, others incorrectly assume that the type of protein (transcription factor) does not influence the outcome, overlooking scenarios where the dosage of a transcription factor is crucial for normal development and function.

**Suggestions for Instructional Improvement:**

- **Enhance Understanding of Gene Dosage:** Emphasize scenarios where the dosage of a gene product is critical, using examples where haploinsufficiency leads to phenotypic changes.
- **Clarify Concepts of Dominance:** Provide clearer explanations and examples of how dominance works, particularly in homozygous organisms, and discuss cases like haploinsufficiency or incomplete dominance where one copy is not sufficient.
- **Discuss Functional Roles of Genes:** Use case studies or examples where the specific roles of genes, such as those encoding transcription factors, critically influence phenotypic outcomes, emphasizing the importance of both the gene's role and its dosage.

Overall, while some students display a good grasp of genetic principles, there is a need to address specific misconceptions and enhance the depth of understanding concerning gene dosage effects, dominance in homozygous organisms, and the functional implications of genes encoding critical proteins like transcription factors.